\documentstyle[preprint,prb,aps]{revtex}
\begin{document}
\draft
\title{Thermomagnetic history effects in SmMn$_2$Ge$_2$}
\author{Sujeet Chaudhary$^1$\footnote{{\it Corresponding Author};\\ E-mail: sujeetc@cat.ernet.in; FAX: +91-731-488300}, M.K. Chattopadhyay$^{1}$, Kanwal Jeet Singh$^{1}$, S. B. Roy$^{1}$, P. Chaddah$^{1}$ and E.V. Sampathkumaran$^{2}$}
\address{$^{1}$Low Temperature Physics Laboratory, Centre for Advanced Technology, Indore 452 013, India}
\address{$^{2}$Tata Institute of Fundamental Research, Homi Bhabha Road, Colaba, Mumbai 400 005, India} 
\date{\today}
\maketitle
\begin{abstract}
The intermetallic compound SmMn$_2$Ge$_2$, displaying multiple magnetic phase transitions, is being investigated in detail for its magnetization behavior near the 145 K first order ferromagnetic to antiferromagnetic transition occuring on cooling, in particular for thermomagnetic history effects in the magnetization data. The most unusual finding is that the thermomagnetic irreversibility, [= M$^{FCW}$(T)-M$^{ZFC}$(T)] at 135 K is higher in intermediate magnetic field strengths. By studying the response of the sample (i.e., thermomagnetic irreversibility and thermal hysteresis) to different histories of application of magnetic field and temperature, we demonstrate how the supercooling and superheating of the metastable magnetic phases across the first order transition at 145 K contribute to overall thermomagnetic irreversibility.
\end{abstract}
\pacs{}
\newpage
\section{Introduction}
There have been a continued interest in the past decade in understanding the magnetization behavior of the body centered tetragonal rare-earth transition-metal germanides and silicides (RMn$_2$Ge$_2$ and RMn$_2$Si$_2$). Various kinds of magnetic phase transitions, viz., paramagnetic (PM) to ferromagnetic (FM), PM to antiferromagnetic (AFM), FM to AFM, AFM to FM or ferrimagnetic state (at low temperature, T) can be observed in different compounds belonging to this class of materials.\cite{1,2,3,4,5,6,7,8,9,10,11,12,13,14,15,16,17} The unit cell in these compounds, consists of layered structure ..-Mn-Mn-R-Ge(or Si)-R-Mn-Mn-.. stacked along the c-axis. Within the ab-plane, the Mn-Mn interaction is FM-like for all temperatures below the highest ordering temperature. It is known that in all the RMn$_2$Ge$_2$ and RMn$_2$Si$_2$ compounds, the various properties including magnetization (M) are strongly dependent on the intralayer Mn-Mn distance, d$_{Mn-Mn}^a$ (Ref.[2]). It has been established that there exist a critical value of d$_{Mn-Mn}^a$ below (above) which the Mn-spins in one FM layer interact antiferromagnetically (ferromagnetically) with those in the neighbouring FM-layer.\cite{2} Among the various members of the entire family, the compound SmMn$_2$Ge$_2$ is unique, in the sense that the intralayer d$_{Mn-Mn}^a$ is very close to the critical value of 2.84A at room temperature.\cite{2,3} Accordingly, significant structural distortions occur in the unit cell of SmMn$_2$Ge$_2$ as one varies temperature in the range of 360 to 5 K.  With a PM to FM transition around 350 K (i.e., T$_C$), these distortions leads to (i) an intermediate-T FM to AFM transition around 145 K (T$_1$), and (ii) a low-T AFM to re-entrant FM transition near 100K (T$_2$).\cite{2,6} In addition, the easy axis in SmMn$_2$Ge$_2$ changes from $<$001$>$ above $\approx$145 K to $<$110$>$ below $\approx$100 K.\cite{2,6} 

The intermediate AFM-regime can undergo metamagnetic transition for magnetic field strengths, H $\approx$ 5 kOe, by which the alternate antiparallel spin-configuration of the FM-layers is transformed to a parallel one.\cite{7,8,9,10} Neutron scattering,\cite{11} NMR \cite{12} and Mossbauer \cite{13} studies have revealed that the magnetic structure of each magnetic phase possessed non-collinear arrangement of spins. Later, a thorough neutron scattering investigation\cite{14} on a sample containing isotopically enriched Sm, revealed the existence of much more complicated cone-structures in AFM as well as low-T re-entrant FM-regime. This ternary compound SmMn$_2$Ge$_2$ has gained further interest due to the observed giant magnetoresistance (GMR) associated with the AFM  phase.\cite{7,8,9,15,16} Magnetoresistance of varying magnitude from 8\% to about 16\% was reported in these studies. Technologically, this is interesting since most GMR materials are artificially grown as thin film multilayers. 

At this juncture, we would like to recall that we reported\cite{18,19,20} the thermomagnetic history effects across the first order FM to AFM transition in polycrystalline Ce(Fe$_{0.96}$Al$_{0.04}$)$_2$. It was shown that the supercooling and superheating of two magnetic phases across the first order transition (FOT) leads to metastable behaviour which resulted in the thermomagnetic irreversibility (TMI) that was found to increase with the increase in H, in sharp contrast with the TMI observed, say for example, in spin-glass\cite{21} and long range magnetically ordered systems\cite{22,23} which gets suppressed with the increase in H. In particular, it was found that both M and magnetoresistivity ($\rho$) at any (T,H) point below the Neel temperature (T$_N$) were strongly history dependent. Although all the generic features associated with the FOT were observed in Ce(Fe$_{0.96}$Al$_{0.04}$)$_2$, the superheating signatures were quite subtle compared with those associated with supercooling.\cite{18,19,20} In this context, the compound SmMn$_2$Ge$_2$ provides an unique opportunity to probe the aspect of TMI in the view of two first order transitions, as it has been established that both the transitions -- at T$_1$, and T$_2$ -- are first order in nature.\cite{6,8,17} This system thus is a natural choice for observing all the characteristics of a FOT (including superheating) as we approach the AFM-regime (displaying negligible moment relative to the FM-state) both while heating from the low-T FM-regime as well as while cooling from the high-T FM-regime. We thus demonstrate through TMI measurements the observation of the metastable phases of both kinds (i.e., supercooled and superheated) so clearly in SmMn$_2$Ge$_2$. In this paper, we present the thermomagnetic history effects observed on variation of field and temperature across the 145 K first order transition in SmMn$_2$Ge$_2$. Thermomagnetic history effects associated with the low temperature transition (i.e., at $\approx$100 K) in this compound are still under investigation, and are not addressed in this paper.
 
\section{Experimental Details}

Polycrystalline samples of SmMn$_2$Ge$_2$ were prepared by argon-arc melting. Details of sample preparation and characterisation can be found in Ref.[15]. The M vs. T and/or H data have been recorded using a commercial SQUID magnetometer (Model MPMS5) with a scan-length of 4 cm. The measurement of M is being done in three different experimental protocols, viz., Zero field cooling (ZFC), Field cooled cooling (FCC) and Field cooled warming (FCW). These protocols are explained in detail in Ref. [19].

\section{Results and Discussion}

Before we present the detailed results on thermomagnetic history effects across T$_1$, we would like to discuss the M vs. T as well as M vs. H behavior, to serve as a prelude to interpret TMI data. Figure 1 shows the M vs. T plot for SmMn$_2$Ge$_2$ sample recorded in a low-field of 50 Oe both in ZFC and FCW protocols. The PM to FM transition takes place at T$_C$ $\approx$345 K. This high T FM-phase continues down to T$_1$ $\approx$145 K, below which the M vs. T curve displays a sudden loss of M thereby entering into an AFM-regime, which persists down to $\approx$100 K. Below this T, which we are referring to as T$_2$, the magnetization once again increases due to the formation of the (low T) re-entrant FM-phase. [Henceforth, these two FM-phases existing at high and low T would be represented by FM1 and FM2, respectively.] It may be noted in passing that the high T FM1-phase (extending over a T-range from 145 to 345 K) displays a concave curvature for T$<$ 180 K, in contrast to a typical M vs. T behavior represented by Brillouin function. A large TMI [i.e., M$^{FCW}$(T) $\neq$ M$^{ZFC}$(T)] is distinctly observed below 100 K. Conventionally, such a TMI in M$^{FCW}$ and M$^{ZFC}$ is commonly taken as finger-print of spin-glass behaviour.\cite{21} However, it is now established that long range magnetically ordered systems (i.e., FM, AFM, etc.) can also show significant TMI in their M vs. T data,\cite{22,23} which arises mainly from hindrance to domain-rotation caused by the magnetocrystalline anisotropy and/or domain-wall pinning effects. One may recall that when TMI arises due to these domain-related effects, it gets suppressed with increase in H. Thus, the small TMI in the FM1-phase of SmMn$_2$Ge$_2$ above 145 K (despite the low field of 50 Oe) is definitely indicative of relatively small domain-wall pinning effects. Besides, in view of highly anisotropic magnetization behavior of SmMn$_2$Ge$_2$, the small TMI in the FM1-phase also suggests that either there is a relatively small magnetocrystalline anisotropy or there is some preferential orientation of $<$001$>$ grains parallel to applied H in this polycrystalline sample. However, on the basis of data of Fig. 1 alone, it is not possible to decipher which one of the above two factors is causing the small TMI observed above T$_1$.

To know about the H-dependence of the TMI behavior in SmMn$_2$Ge$_2$, we show in Fig. 2 the M vs. T plots for H, namely 5 kOe, 20 kOe and 50 kOe. We find that, \\ 
\begin{enumerate} \item Instead of loss of magnetization as observed at low-H (see Fig. 1) in the AFM-regime, a "dip" like feature is now observed in M vs. T plot for H=5 kOe, indicating that the AFM-regime is narrowed down in high H. We also want to draw the reader's attention to the fact that relative to the TMI in small H (see Fig. 1), the increase of H to a moderate value, for instance, 5 kOe has almost smeared out the T$_2$-transition, whereas the T$_1$-transition is less affected qualitatively. Furthermore, in presence of high fields (H $\ge$ 5 kOe), the moment in the FM1-phase is distinctly larger than that in FM2-phase, which is in sharp contrast with the situation at 50 Oe (see Fig. 1). At further higher H, the M vs. T plots do not show any dip in magnetization at T$_2$-transition (i.e., for both H=20 kOe and 50 kOe). Instead, magnetization (in H=50 kOe) rises with T right from T $>$ 30 K. 
\item The TMI (see, e.g., magnetization at 5 K) decreases as one goes from 50 Oe (Fig. 1) to 20 kOe (Fig. 2). However, at 50 kOe, we find that TMI has a different sign, i.e., M$^{FCW}$(T) $<$ M$^{ZFC}$(T). This change in sign of TMI at higher H is quite {\it anomalous}, the origin of which is unclear at present.
\item The peak in M (observed in both FCW as well as ZFC protocols) shifts to a lower T with increase in strength of H, and this implies complex intermediate tempearture magnetic phase at higher H. 
\end{enumerate}

As mentioned in the preceeding discussion, in order to identify these critical fields of metamagnetic transition H$^{meta}$(T), we recorded various isothermal M vs. H plots within the AFM-regime. Figure 3a shows one such plot at 135 K, in which (a) the initial M vs. H curve recorded after reaching to 135 K point strictly in ZFC-manner from above T$_C$ (i.e., the {\it virgin-curve}), (b) a portion of H-reversal cycle from the maximum 50 kOe  down to -50 kOe through 0 T (i.e., {\it reverse-envelope curve}), and (c) a portion of field ascending M-H curve initiated after field excursion from -50 kOe, i.e., {\it forward-envelope curve}) are shown for the SmMn$_2$Ge$_2$ sample. 

A slow increase in M until about 3 kOe is consistent with the low-field AFM-state in the present SmMn$_2$Ge$_2$ sample at 135 K. With further increase in H, a number of jumps in M  are clearly resolved until about 6 kOe along the virgin M-H branch (Fig. 3a). At higher field strengths, the observation of usual saturation-like behavior (typical of a FM-state) indicates the completion of field-induced AFM to FM transition in SmMn$_2$Ge$_2$ sample. The observed randomness in both the magnitude as well as the position of these various jumps (which are as much as 10 in number in the present case of SmMn$_2$Ge$_2$ sample (see Fig. 3a)), may be attributed to a distribution of H$^{meta}$ due to possible inhomogenieties/disorder in the polycrystalline sample together with the high anisotropy of SmMn$_2$Ge$_2$. From the first and the last jumps in the virgin M vs. H curve, one can identify the critical field for the onset and completion of metamagnetic transition at 135 K. In Fig. 3b, we plot these two critical fields as a function of T covering the entire AFM-regime. We shall refer to this phase diagram when we discuss our measurements to look for metastable (supercooled/superheated) states by varying T in fixed H. Finally, it is worth noting that the virgin M vs. H curve lies anomalously outside the full hysteretic loop obtained by cycling the field between +50 kOe and -50 kOe (see Fig. 3a). A similar anomalous virgin curve was observed in the M vs. H and $\rho$ vs. H data of Ce(Fe$_{0.96}$Al$_{0.04}$)$_2$.\cite{18,19,20} 

We stress here that the M$^{FCW}$(T) data presented in Fig. 1 and Fig. 2 have been recorded after cooling the sample (in presence of respective H's) across the {\it two} FOTs (and not one FOT as reported in Ref.[18,19,20] in Ce(Fe$_{0.96}$Al$_{0.04}$)$_2$ system). We further point out here that a great care is required when dealing systems like the present SmMn$_2$Ge$_2$ sample, since metastabilities across both the first order transitions (i.e., FM1 to AFM, and AFM to FM2) may mask the magnetic character of that particular phase (due to the supercooling and superheating of various phases across the two transitions.)\cite{24} Thus, the TMI effects in SmMn$_2$Ge$_2$ may be different not only from those encountered in only FM- or only AFM-ordered compounds,\cite{22,23} but also from the TMI observed across a single FOT from FM to AFM as discussed in Ref.[18,19,20]. In this paper, we will focus on TMI effects only across the T$_1$-transition. To limit the contributions to the overall TMI (say near the transition-temperature) arising from the metastable effects related with domain pinning/hindrance, the strength of H should be small enough so as not to drive the AFM-state into a FM-state while preparing the field-cooled state from, say at 135 K.

In Fig. 4, we show the effect of strength of H on the TMI (near the T$_1$-transition) obtained from the M vs. T plots which have been recorded while warming the sample; \\
(a) after cooling in zero field from above T$_C$ to 120 K (thereby ensuring the initial phase at 120 K to be purely AFM) at which the appropriate field is applied (i.e., ZFC protocol), and \\
(b) from 135 K which the sample reached in presence of H when cooled from a temperature T$_0$ ($>$ T$_1$) (i.e., FCW protocol). [The T$_0$ is the temperature upto which the sample is warmed while recording M$^{ZFC}$(T)-data as explained in step-(a) above.]\\  

Note that the M vs. T plots in Figs. 4a-g are normalized to their maximum value at the transition T$_1$, to allow a proper comparison of the effect of the strength of applied field on the TMI. It can be seen that the TMI observed below T$_1$ [i.e., M$^{FCW}$(135 K)- M$^{ZFC}$(135 K)] rises with the increase in H from 20 Oe to 2 kOe (see Fig. 4h). As discussed above, such a TMI between ZFC and FCW -- i.e., increasing with field -- indicates (at first place) the first order nature of the magnetic transition taking place at T$_1$, rather than having an origin due to domain-related behavior wherein the TMI gets suppressed with the increase in H.\cite{22,23} However, with further increase in H, the drastic decrease of TMI at 135 K is observed. It thus turns out from the foregoing data that TMI at intermediate H (i.e., $\approx$ 2-4 kOe) is higher than the TMI observed both in low (i.e., 20 Oe and 1 kOe) or high fields (i.e., 6, 10, and 20 kOe). This is remarkably a peculiar finding. We now discuss the possible origin of this behavior.

We note from Fig. 3b, that the sample undergoes a complete metamagnetic transition at 120 K by $\approx$ 7.5 kOe field (see Fig. 3b), with the result that along the ZFC warming M vs. T curve (open data symbols in Figs. 4a-g), the sample is completely in FM-state for H = 10 kOe and 20 kOe, and partly in AFM- and partly in FM-state for 3.2 kOe $<$ H $<$ 7.5 kOe. The TMI increasing with the increase in H from 20 Oe to 2 kOe should predominantly be due to the superheating of the (metastable) AFM-phase while warming along the ZFC-protocol and also due to the additional metastability in M$^{FCW}$(T) along the FCW-run (filled  data symbols in Figs. 4a-g) as there is always a probability that a finite fraction of the FM1-phase may get supercooled down to 135 K (the starting T of the FCW-run). This initial trend of TMI [i.e., increasing with H, and consistent with the arguments by Chaddah and Roy \cite{25}] is also identical to the one observed in the M(T,H)- and $\rho$(T,H)-data of Ce(Fe$_{0.96}$Al$_{0.04}$)$_2$ exhibiting a FOT from FM to AFM at $\approx$100 K. \cite{18,19,20} The TMI in SmMn$_2$Ge$_2$ for H $\approx$4 kOe or higher may be attributed to  arise from both the metastable-effects associated with domain-related causes, as well as the metastable-effects associated with the FOT at T$_1$. This is so, because for H $\ge$4  kOe (but $\le$7.5 kOe), the sample also consists of a finite fraction of FM1-phase (see Fig. 3b) along the ZFC-cycle right from 120 K onwards, which would result in higher M$^{ZFC}$ below $\approx$T$_1$, thereby resulting in a small TMI as is experimentally observed in Figs. 4d and 4e. This finite FM-fraction in ZFC would also contribute in further reduction in TMI with increase in H, becasue of the known supression of TMI arising due to domain-related effects with increase in H.\cite{22,23} On the other hand, given the first order nature of the transition at T$_1$, one may still argue that the FCW data of M vs. T plot (for 10 kOe or higher) could still result due to the supercooling of high-T FM1-phase. We shall defer this question for time being, and point out that (a) the thermal hysteresis at the onset of transition while cooling and warming in presence of field,\cite{26} and (b) its dependence on H are both instructive to know the dominance of metastable effects associated with any FOT. In the next section, we shall now present results of H-dependence of thermal hysteresis across the T$_1$-transition.

In Fig. 5, we show the results of thermal cycling of SmMn$_2$Ge$_2$ sample across the T$_1$-transition in presence of different field strengths. (It should once again be noted that the M vs. T data for different H's (for both FCC- as well as FCW- protocols) is normalised with respect to the highest M-value observed along the FCC-curve). For each H, we first brought the sample to 150 K in ZFC-manner from above T$_C$, then the field is applied and FCC-data is first collected down to 120 K followed by recording of FCW-data by warming the sample upto and above T$_1$. Akin to the multiple jumps seen in the M vs. H curves within the AFM regime (Fig. 3a), more than one jump in M are also observed in these thermal cyclings as well. A significant amount of thermal hysteresis is visible for all the field strengths upto H=4 kOe. [This presence of thermal hysteresis\cite{26} itself indicates the first order nature of T$_1$-transition.] However, the absence of thermal hysteresis between the FCC- and FCW-curves recorded in high magnetic field strengths (see inset to Fig. 5 for H=20 kOe case) immediately confirms that T$_1$-transition is a second-order transition in presence of higher H. In our opinion, this transition at H=20 kOe involves a gradual transformation of FM1-state with a particular easy-axis to another FM-state (possibly FM2-like) with different easy-axis. (i.e., the transition in higher H is associated with the change in anisotropy in SmMn$_2$Ge$_2$ near 145 K.)

The amount of thermal-hysteresis (near the mid-point of the total magnetization-change at T$_1$-transition) increases from about $\approx$7 K in 20 Oe, and to $\approx$12 K at 0.5 kOe to $\approx$13 K at 1 kOe, and then decreases to $\approx$10 K at 2 kOe, and then to $\approx$4 K at 4 kOe. The initial increase of thermal hysteresis from 20 Oe to 1 kOe is consistent within the picture of FOT.\cite{25} This is explained as follows: while cooling (warming) the sample in presence of H from 150 K (120 K) a finite fraction of high (intermediate) temperature FM1 (AFM) phase supercools (superheats) below (above) T$_1$ down (up) to the lower-T (higher-T) metastable limit T$^*$(H) (T$^{**}$(H)), which is the temperature at which the hysteresis collapses on lower-T (higher-T) side.\cite{24} Within this FOT picture, it is very well necessary that the hysteretic regime in SmMn$_2$Ge$_2$ should also widen with increase in H. The results shown in Fig. 5 support this beyond any doubt until 1 kOe. Although further increase in H suppresses the hysteresis, the presence of hysteresis above H $\ge$ 1 kOe itself is indicative of first-order like transition upto 4  kOe.\cite{26} (or upto the H, where one could still observe the hysteresis in SmMn$_2$Ge$_2$.) In field strengths H $\ge$1 kOe, the reduction in the thermal-hysteresis with H could be associated with the varying fractions of AFM- and FM- phases in SmMn$_2$Ge$_2$ while traversing the phase coexistence regime (see Fig. 3b) on either sides (during the FCC- and FCW-runs) or to the distribution in H$^{meta}$(T) in the sample.

\section{Conclusion}     
We have probed the thermomagnetic history effects for the first time in a compound, viz., SmMn$_2$Ge$_2$, exhibiting two first order magnetic transitions. We have mainly focussed  the present study on a FM to AFM transition occuring around 145 K in SmMn$_2$Ge$_2$. The most unusual finding is that higher TMI is observed at intermediate field strengths. The results reveal that there are two kinds of metastable effects giving rise to the observed TMI in SmMn$_2$Ge$_2$: (a) the metastable effects associated with a first order transition (i.e., supercooling and superheating) dominate at lower fields, and (b) the metastable effects resulting from the hindrance to the domain rotation process caused due to the high magnetocrystalline anisotropy and/or due to the pinning of domain-walls at lattice defects dominate above 4 kOe.

\begin{figure}
\caption{M vs. T plot of SmMn$_2$Ge$_2$ sample in presence of 50 Oe field recorded in ZFC (open triangle symbols) and FCW (filled triangle symbols) protocols covering the 4.5 to 360 K T-range. The different magnetically ordered phases are labelled as FM1, AFM and FM2 in their respective T-regimes. Also marked in figure are the transitions, viz., T$_C$, T$_1$ and T$_2$ separating these different magnetic phases. See text for more details.}
\caption{Effect of magnetic-field strength on the thermomagnetic irreversibility observed between the ZFC- (open symbols) and FCW- (filled symbols) magnetization data  recorded between 4.5-360 K in SmMn$_2$Ge$_2$; H=5 kOe (down-triangle symbols), 20 kOe (circle symbols) and 50 kOe (up-triangle symbols).}
\caption{(a)The M vs. H curve at 135 K (which is reached in zero field on cooling from above T$_C$) showing the hysteresis between the forward- and reverse-envelope cycles recorded between H=+50 kOe and H=-50 kOe. Note that the virgin magnetization branch lie outside the complete hysteretic-loop. The rise in M at $\approx$3 kOe is due to the field-induced AFM to FM transition which takes place through successive random jumps in M until it is finally completed at $\approx$6 kOe. See text for further details. (b) The magnetic phase diagram of the investigated SmMn$_2$Ge$_2$ sample highlighting the completely AFM-regime separated from the completely FM-regime through a mixed phase regime (i.e., comprising of AFM and FM fractions). The various points along the two boundaries on either side of this mixed-phase regime are obtained by inferring the onset (square symbols) and completion (triangle symbols) of metamagnetic (i.e., AFM to FM) transition at different temperatures within the AFM-regime (i.e., from various M-H loops like e.g., Fig. 3a for T=135 K).}
\caption{(a)-(g) The effect of field strength on the TMI between the ZFC (open symbols) and FCW (filled symbols) M vs. T runs. While ZFC runs for each field are initiated from 120 K which the sample reached in zero field on cooling from above T$_C$, the FCW-runs have been initiated from 135 K which the smaple reached when carefully cooled after the completion of the ZFC-run to more than 200 K; (a) H=20 Oe, (b) 1 kOe, (c) 2 kOe, (d) 4 kOe, (e) 6 kOe, (f) 10 kOe, and (g) 20 kOe. Note that in order to have a proper comparison, the magnetization-data is normalised with respect to the peak-value observed near the T$_1$-transition. (h) The thermomagnetic irreversibility at 135K [i.e., M$^{norm, FCW}$(135 K,H) - M$^{norm,ZFC}$(135 K,H)] as a function of applied field strength.}
\caption{The normalised magnetic-moment as a function of temperature for SmMn$_2$Ge$_2$ while cooling in presence of field from 150 K down to a temperature between 115-125 K (i.e., FCC-protocol), and subsequent warming to above 150 K (i.e., FCW-protocol). Main Panel: H=20 Oe (diamond symbols), 0.5 kOe (down-triangle symbols), 1 kOe (square symbols), 2 kOe (up-triangle symbols) and 4 kOe (circle symbols). The open and filled symbols respectively represent the FCC- and FCW-data. The inset shows the thermal cycling (i.e., FCC and FCW runs) for H=20 kOe. Within the error in temperature-measurements ($\le$0.5 K), there is no hysteresis in FCC- and FCW-data for this high field strength (i.e., 20 kOe).}
\end{figure} 

\end{document}